# Information content of contact-pattern representations and predictability of epidemic outbreaks

**Petter Holme**

Department of Energy Science, Sungkyunkwan University, 440-746 Suwon, Korea
IceLab, Department of Physics, Umeå University, 90187 Umeå, Sweden
Department of Sociology, Stockholm University, 10961 Stockholm, Sweden

To understand the contact patterns of a population—who is in contact with whom, and when the contacts happen—is crucial for modeling outbreaks of infectious disease. Traditional theoretical epidemiology assumes that any individual can meet any with equal probability. A more modern approach, network epidemiology, assumes people are connected into a static network over which the disease spreads. Newer yet, temporal network epidemiology, includes the time in the contact representations. In this paper, we investigate the effect of these successive inclusions of more information. Using empirical proximity data, we study both outbreak sizes from unknown sources, and from known states of ongoing outbreaks. In the first case, there are large differences going from a fully mixed simulation to a network, and from a network to a temporal network. In the second case, differences are smaller. We interpret these observations in terms of the temporal network structure of the data sets. For example, a fast overturn of nodes and links seem to make the temporal information more important.

## Introduction

**Background and problem statement**

Epidemics of infectious disease are complex phenomena involving processes at different scales. The smallest and fastest processes happen in the bodies of the infected people, as the pathogen enters and colonize the host. The largest and slowest processes involve the evolution (or rather co-evolution) of the pathogen and hosts and the development of new treatments. Intermediate to these are the movements of infected hosts and the response of society to an emergent outbreak. In this work, we consider the problem of predicting an outbreak based on (partial) knowledge of the small and intermediate scale processes, while treating evolutionary processes as constant. In other words, we assume that we have some understanding of the pathogen (and pathogenesis), the evolving outbreak, and the way people move and come in contact in such a way that the disease can spread. This is the type of challenge modelers face when there is an outbreak of a new pathogen, or a pathogen in a previously unaffected population [1,2].

To be more specific, we assume the small-scale processes are well modeled by a compartmental model—a scheme dividing the population into categories with respect to the disease and assigning transition rules between the classes. We will focus on the Susceptible-Infectious-Recovered (SIR) model—the canonical model of diseases that makes infected persons immune upon recovery [2]. (For details about the simulation, see the *Methods* section.) A compartmental model needs to be complemented by a model of how people meet and interact, i.e. the contact patterns. We will compare three ways, or levels, of including contact information. The first way is to assume that we know, or are able to model, who is in contact with whom, and also the time of the contacts. We refer to this level of capturing the contact structure as a *temporal network* representation [3-5]. The next level is a *static network* [6–8] representation where we assume that we know who that can be in contact with whom, but nothing about when the contacts happen. The last level, with the least information content, is a *fully mixed* case [2] where everyone is equally likely to be in contact with anyone else at any time. The three levels of information content are illustrated in Fig. 1. The question of this paper is how including more in-formation about the contacts—going from the well mixed, to the static, and then to the temporal network representation—changes different aspects of our ability to predict the final outcome of the epidemics. We investigate not only the predicted final outbreak size given there is an outbreak from an unknown node, but also given the current state of an ongoing outbreak at a specific time (the *breaking time*) [9], i.e. assuming more knowledge. The histogram of final outbreak sizes defines what we call unpredictability, or outbreak diversity. There are of course many ways of conceptualizing predictability and defining measures for it, but a broad histogram of outbreak sizes means that there is an inherent stochasticity in the outbreaks that makes it hard to predict the final outcome.

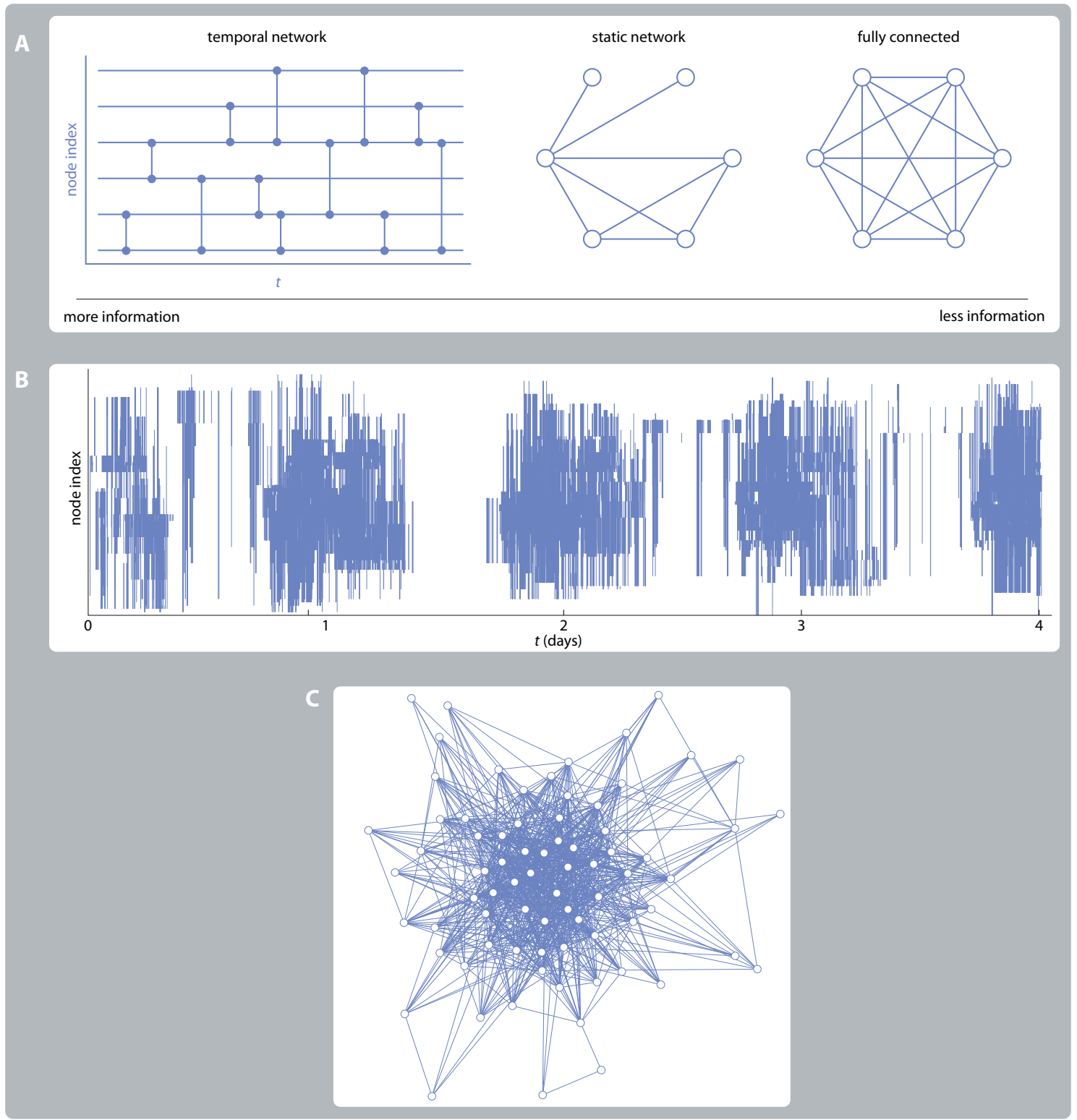

**Fig. 1. Illustration of the representations of contact patterns containing different levels of information.** (A) shows the three levels, with respect to information content, of contact representations. The temporal network is visualized using a time line of nodes. If two nodes are in contact at some time, then there is a vertical line at that time. (B) gives a real world example of the time-line plot of the temporal network representation of the *Hospital* data. The horizontal lines of (A) are, however, omitted. The indices are chosen to minimize the total length of vertical lines. (We chose the *Hospital* data as an example because it has very clear temporal structures, with a conspicuous diurnal pattern.) (C) shows the corresponding projected static network of node pairs with more than five contacts.

We use empirical contact sequences as the underlying contacts structure for our simulations. We include them either as they are (like temporal networks), or reduce them to static networks, or to fully mixed models. When we project out information, we attempt to keep as much information as possible (cf. Ref. [10]—so the fully mixed models keep the overall contact frequency of the original data and the same number of contacts will take place over the static networks as in the original data). Then we scan the entire parameter space of the SIR model on these three representations of the contact patterns.

**Previous studies**

The fully mixed case is by far the most studied disease-spreading framework. There are several textbooks and review papers discussing their use. We recommend Ref. [2] as an introduction. If one has no information about human contact patterns, we have to treat each agent the same, which leads to the fully mixed approach. But on the other hand, we almost always do have more information—for example, we know that there is a broad distribution of the number of sexual partners, which affects the spread of e.g. HIV [11]. The static network paradigm has been around for at least two decades and profoundly influenced theoretical epidemiology [8]. In addition to making predictions more accurate, one major contribution has been to put an emphasis on the different roles and importance of individuals. Networks provide a framework to explain phenomena such as super spreaders or to find the optimal set of people to vaccinate or quarantine [8]. Temporal network epidemiology [4] is the youngest of these branches of theoretical epidemiology as categorized by their representation of contact patterns. There are several studies showing that including time in the contact representations does make a big difference [12–14] in the predicting outbreak sizes. One conclusion is that bursty time series of contacts slow down spreading processes [14]. Another observation is that the birth and death of nodes and links are even more important for disease spreading [15]. A few studies investigate how to exploit the temporal structure to mitigate outbreaks [16,17].

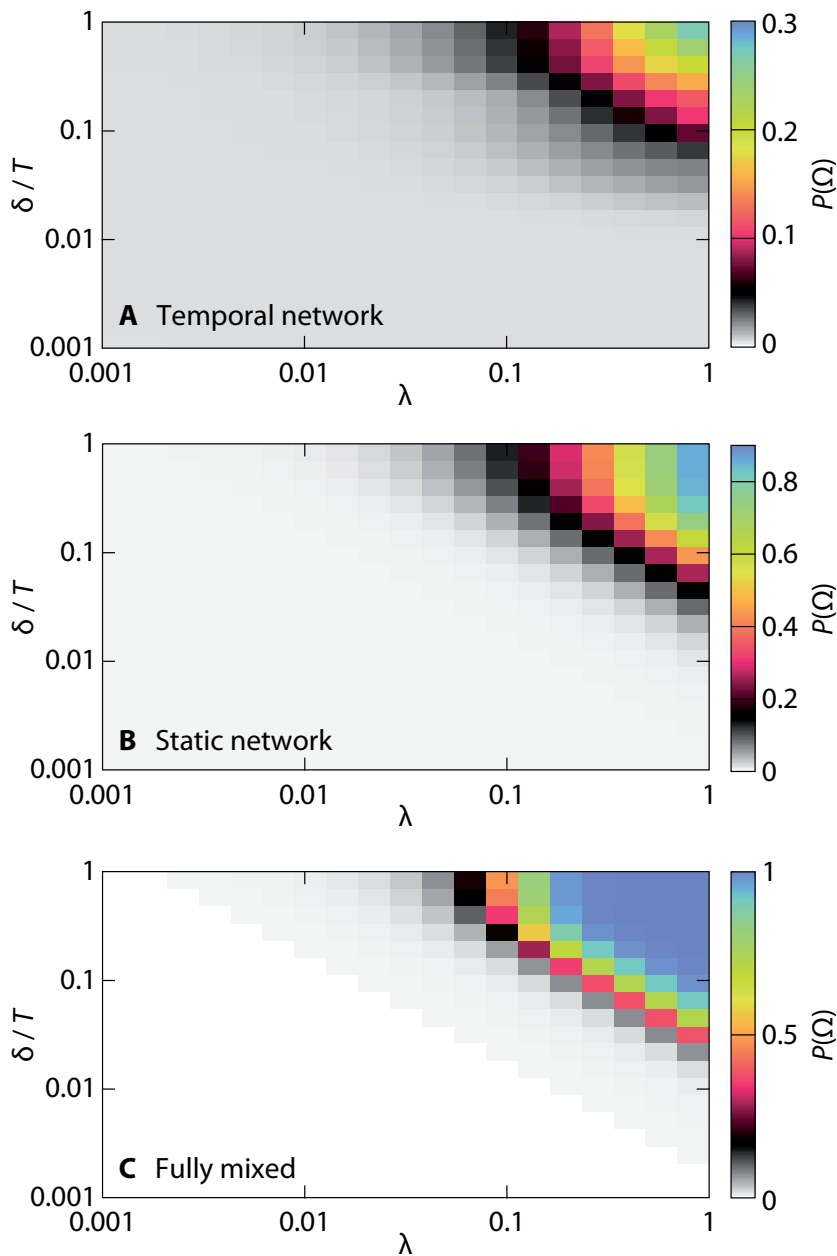

**Fig. 2. Outbreak sizes for the three representations of contact patterns for the *Gallery* data.** (A) shows the fraction of recovered individuals at the end of the outbreak for the temporal network data. (B) shows the corresponding plot for the static network of aggregated contacts. (C) is the plot for the fully connected network. The corresponding plots for the other data sets can be found in Supplementary Fig. S1.

Another line of research recognizes that the contacts are not independent of the disease itself. People would change their contact patterns if they become infected or perhaps just from awareness of the outbreak [18]. In our work, we ignore to model this effect and focus on the impact of the structure of the contact patterns *per se*.

The previous work most similar to ours is probably Ref. [9] where we look at the decay of unpredictability (defined in a similar way to this paper) as a function of time. In that paper, we investigated (using network models) how static network topology influences this decay.

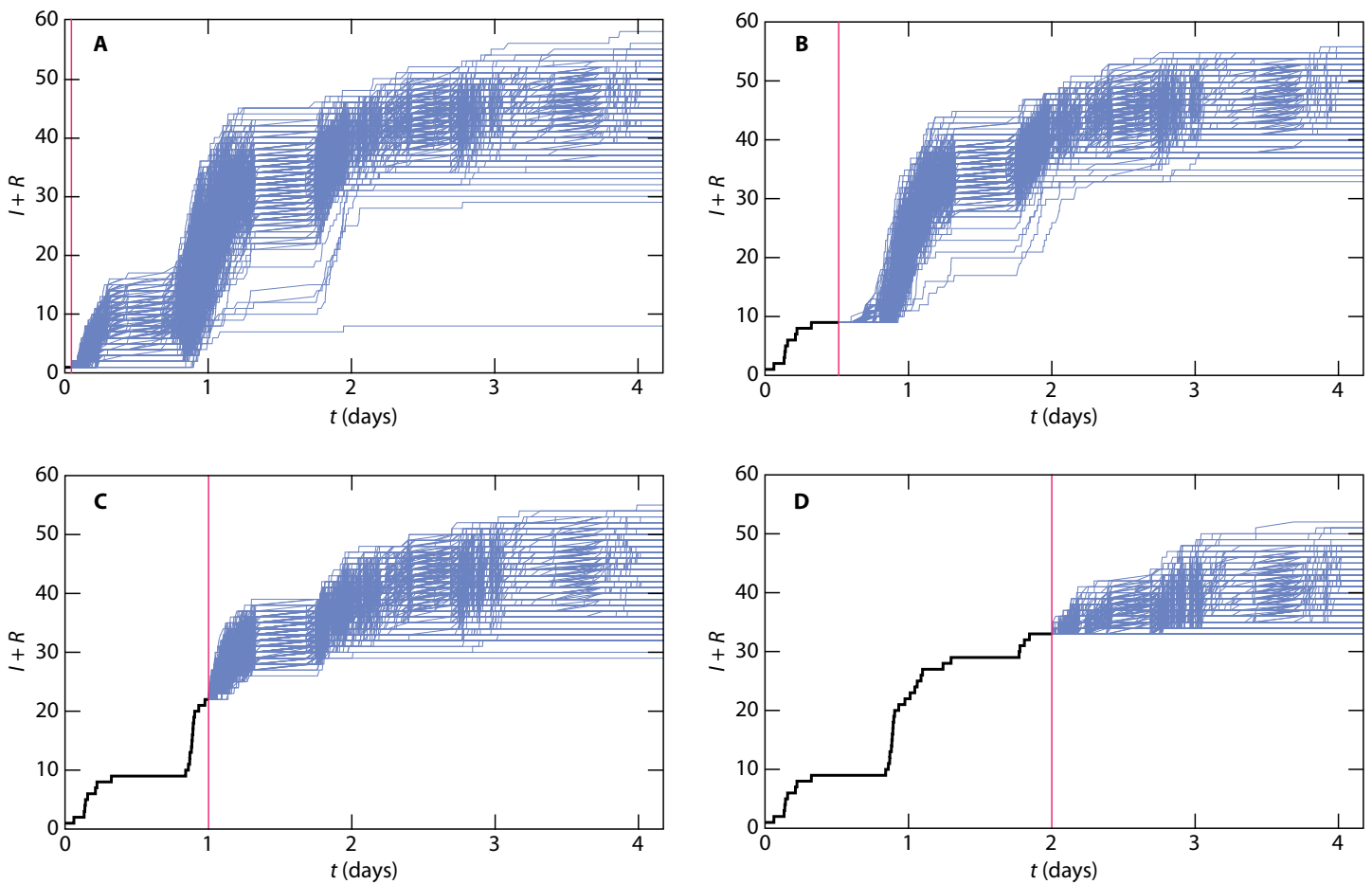

**Fig. 3. Example of continuations of outbreak trajectories given the state of an outbreak at certain breaking times *t*.** This figure illustrates our method to measure the deviation from a predicted outbreak size. The thin lines show 1000 possible future trajectories from the breaking point (indicated by the horizontal line). The thick line shows the trajectory actually taken up to the breaking time. The simulations are from the temporal network representation of the *Hospital* data with parameter values $\delta$ = 0.6 and $\lambda$ = 0.1.

**Empirical data**

Our starting point is empirical data sets of temporal human proximity networks—records of two persons being close to each other, and when these contacts happen. Any non-vector-borne infection whose pathogens cannot survive for extended periods outside a host do spread over proximity networks. However, the exact requirement for two persons to be close enough, and the exposure to be long enough, for the disease to spread varies for different diseases. Human proximity data is, however, hard to obtain at a resolution enough to model the epidemics of a specific disease. Instead of focusing on a particular disease (i.e. fine tuning the SIR parameter values to this disease), we scan the entire parameter space and thereby study features general to all SIR type diseases. We list

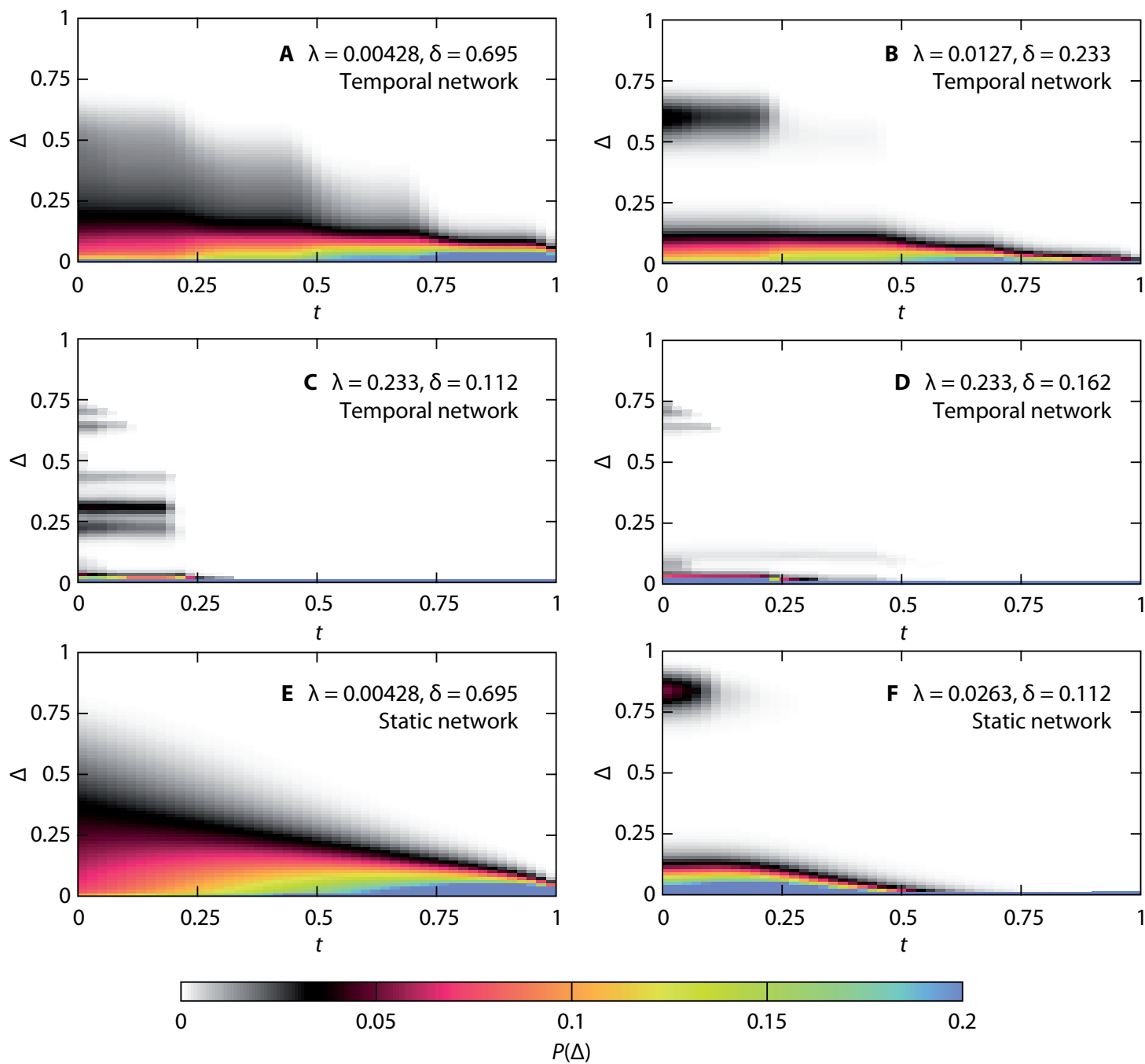

**Fig. 4. Time evolution of deviation from other outbreaks (*Hospital* data).** Panels A–D shows data from the temporal networks while E and F are for the static networks. A shows a typical plot for low-$\lambda$ and large-$\delta$ values. B shows a bimodal histogram for intermediate $\lambda$. D and E represent large-$\lambda$ and low-$\delta$. Although the change in $\delta$ is not that large between D and E, the pattern of the deviations is. E shows the a plot for the static network representation with the same parameter values as panel A. F shows a typical bimodal configuration for the static network case (corresponding to panel B, but for slightly different parameter values).

the sizes, sampling durations, etc., of the data sets in Table 1.

One of our data sets comes from the Reality mining study [19] (*Reality*). In this data set, contacts within a cohort of university students were recorded by the Bluetooth devices of their smartphones. The range of such devices is between 10 and 15 meters. To be able to compare our results to other studies, we use a reduced set of contacts from this data set—the same as in Refs. [15,20].

Another group of proximity data comes from the Sociopatterns project (sociopatterns.org). These data sets are gathered from groups of people wearing radio-frequency identification sensors. Such devices record a contact if two sensors are no further than 1–1.5 m, and the wearers are facing each other. One of these datasets come from the attendees of a conference [21] (*Conference*), another from a school (*School*) [22], another from a hospital (*Hospital*) [23] and yet another from visitors to a gallery (*Gallery*) [24]. The *Gallery* data set comprises 69 days where we use the first three. *School* covers two days and we use both.

A different type of proximity data set that we also study comes from self-reported sexual contacts between female prostitutes and male sex buyers [25]. We call this data set *Prostitution*.

## Results

Now we turn to the results of our analyses. For every data set, our raw output is a four dimensional array of values—a histogram of outbreak sizes as a function of the breaking time and the two parameter values of the SIR model. Of course we need to simplify this output by projecting out different dimensions. For further simplification, we will mostly present the results for one data set in the paper and leave the rest for the supplementary information. Which data set that we chose depends on what feature we will highlight in the discussion. In other words, we do not try to show representative results (which is anyway hard to do objectively), but those that help the discussion of the features of our data collection.

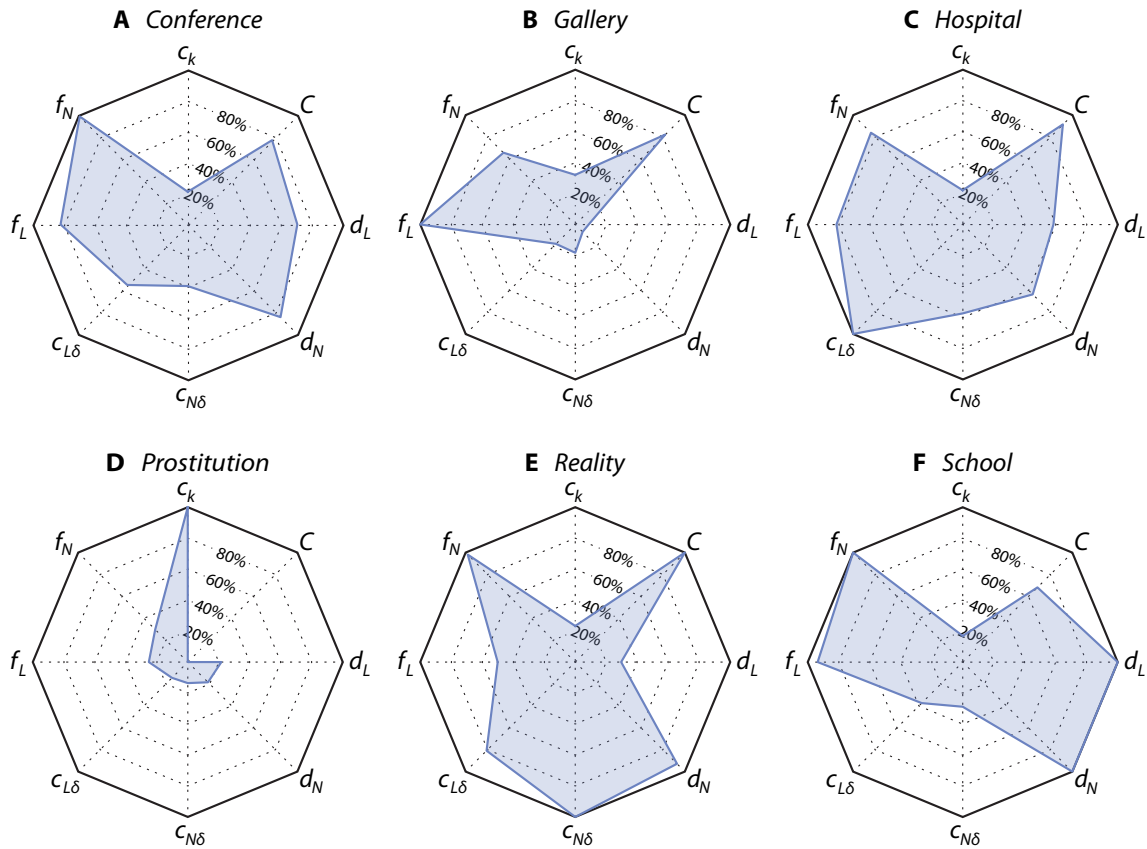

**Fig. 5. Radar plots summarizing the temporal network structures.** The radial component of the areas gives the relative value of the quantity compared to the maximum in among the six data sets. The eight quantities are explained in detail in the *Methods* section. They (and their maximal values) are as follows: $c_k$—coefficient of variation of the degree distribution (maximum 2.24 for *Prostitution*); $C$—clustering coefficient (maximum 0.644 for *Reality*); $d_L$—average duration of links (maximum $0.404T$ for *School*); $d_N$—average duration of nodes (maximum $0.938T$ for *School*); $c_{N\delta}$—node burstiness (maximum 11.1 for *Reality*); $c_{L\delta}$—link burstiness (maximum 15.8 for *Hospital*); $f_L$—fraction of links present at $T/2$ (maximum 0.783 for *Gallery*); $f_N$—fraction of nodes present at $T/2$ (maximum 0.987 for *Conference*).

**Predicting outbreak sizes with no knowledge about who is infected**

In this section, we investigate how the three levels of representations affect the predicted outbreak sizes given no knowledge about who is infected. This type of comparative study has been done previously to show the effects of including (static) network information [1,7,26] and temporal information [13–17]. However, to our knowledge, this is the first time all three levels of representations are considered simultaneously.

In Fig. 2, we plot the average fraction of nodes that are infected during the outbreak for the *Gallery* data set. Both the static network representation and the fully connected picture are rather different from the results for the temporal networks. Briefly stated, without the temporal component, the simulations overestimate the outbreak sizes. One factor is of course the reduced reachability [3] in the temporal networks—the fact that you cannot reach every individual from every other, even though a path in the static network of aggregated contacts could connect them. This effect is more than a question of the existence of such time-respecting paths—assume the existence of such path from *i* to an important spreader *j* hinges on one contact, then chances are high that the outbreak will not reach *j* from *i*, and if the important node *j* is not infected this might reduce the average outbreak sizes much.

The difference between the fully connected and static network is a little bit smaller. Still, among the data sets we test, *Gallery*'s results are most affected by the static network structure. Other datasets, such as *Hospital*, *Conference* and *School*, are more densely connected and thus the spreading is faster and, in effect, more similar to the fully mixed case (cf. Fig. 1C). The static *Gallery* network is stretched out as an effect of time—visitors come, spend some time at the gallery and leave so early visitors would not meet late visitors. In the large-$N$ limit, the static network structure can make a huge difference (the vanishing epidemic threshold for scale-free networks to mention one example [8]). With an exception for the *Prostitution* data, we draw similar conclusions from the other data sets (see Supplementary Fig. S1)—first, the temporal structure makes a larger difference than that the static network structure; second, including this structure makes the outbreaks smaller.

**Outbreak diversity and the approach to high predictability with knowledge about who is infected**

The average value of the outbreak sizes is of course only one type of result that epidemic simulations can give. They can also predict dynamic quantities such as the early inci-dence (number of new cases per time) [27] or the extinction time [28,29], and also distributions of outbreak sizes and times. In this work, we will look further at the distribution of predicted outbreak sizes assuming that we know the state of the outbreak—who is susceptible, infectious or recovered and when the infectious people were infected—at any time $t$. Given this information, the SIR model gives a distribution of outbreak sizes. Once an individual is infected, its final contribution to the outbreak size is determined. Thus, as $t$ increases, the distribution will gradually gather all its weight at zero. How this approach to having total predictability unfolds can vary much. In Fig. 3, we illustrate our method to investigate this. In parallel to a *master run* of the outbreak simulation (the thick line in the plots), we use the state of the system as the seed for $10^3$ independent *auxiliary runs*. From an auxiliary run $i$ starting at the breaking time, we measure the fraction of individuals eventually infected $\Omega_i$. Then we measure histograms of $\Delta\Omega = |\Omega_i - \Omega_j|$ over all pairs of auxiliary runs $i \neq j$, and $10^4$ master sequences.

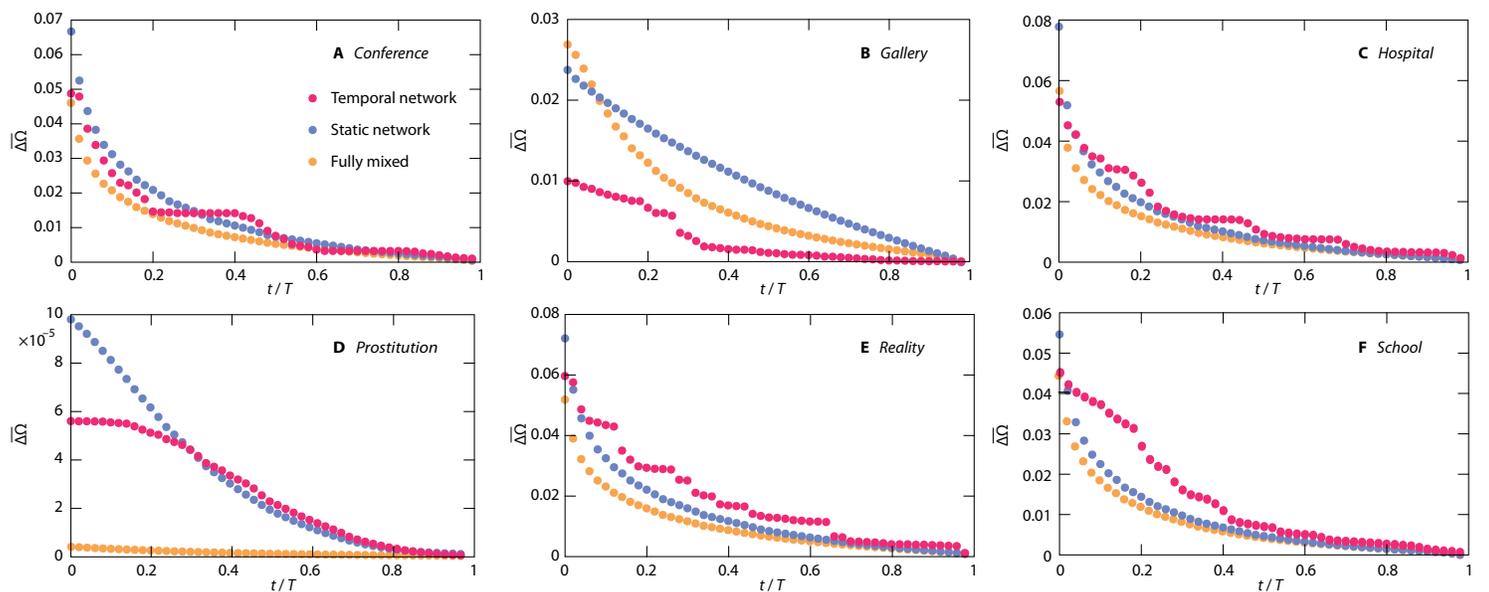

**Fig. 6. The time evolution of the average outbreak diversity.** We investigate the average deviation of pairs of outbreak sizes (given the state of the system at time *t*). Here we show the results for *Gallery* (panels A, B and C) and *Hospital* (D, E and F). For temporal (A and D) and static (B and E) networks and a fully mixed case (C and F).

In Fig. 4, we show a number of examples of $\Delta\Omega$ histograms as functions of time (all from the *Hospital* data set). For the temporal network data, there is a fundamental difference between large $\delta$ and small $\lambda$ on one hand, and large $\lambda$ and small $\delta$ on the other. In the former case (Fig. 4A being a typical example), the deviations are continuously decaying (with the peak always around zero). The decay accelerates with $t$. As $\delta$ decreases, or $\lambda$ increases, the histogram loses its unimodal shape. Typically it turns into a bimodal distribution as seen in Fig. 4B. If $\delta$ decreases, or $\lambda$ increases, further, then the bimodal distribution will split into more, and more well defined, peaks Fig. 4C. This situation, however, exists only in a small fraction of the parameter space. In Fig. 4C, $\lambda = 0.23$ and $\delta = 0.11$. For the next $\delta$-value we measure ($\delta = 0.16$), several of the largest peaks are gone (while some others remain almost identical).

Our interpretation of the above observations is that, if the transmission probability is very small, there is a fairly constant chance for the outbreak to die out. If it was exactly constant, it would give an exponential distribution of extinction times (and probably of $\Delta\Omega$ too). This has been observed before for the SIR model on static networks [29,30] (it is indeed true for our simulations too—see Fig. 4E). The situation for higher transmission probability can be described as a transient when the outbreak can either die or spread. Once it takes off, it behaves rather deterministically [31] (at least in the limit of large population size). This situation can be seen in Fig. 3E, representative of the static networks and fully mixed simulations. This process results in a bimodal distribution. Increasing the transmission probability further while lowering the disease duration makes the process yet more deterministic. At the same time, it also reduces the number of possible outbreak trees. The question, in this parameter region, is not whether there will be an outbreak or not, but which one of a few possible outbreak scenarios that will happen. These few possibilities shows as peaks (or rather lines) in Figs. 4C and D.

**Interlude: temporal network structure of the data sets**

The main theme of this article is to understand the effects of the level of information content of the contact representation on the deviations from the predicted outbreak sizes. For the discussion in the next section, however, we will need a bit more nuanced picture of the temporal network structure of the data sets. Here we examine three different classes of measures of temporal network structure—those characterizing the static network, those characterizing the time series of contacts of individuals and pairs of individuals, and finally those characterizing long-term trends in the activity in the data set.

To summarize the static network structure, the first quantity we study is the coefficient of variation of the degree distribution $c_k$ (the frequency distribution of the number of others the individual interacts with at least once during the sampling period). It is known that a heterogeneous degree distribution makes the spreading faster and further reaching [8,32]. The coefficient of variation is a dimensionless measure of the heterogeneity of a distribution.

Another static network measure that is known to affect disease spreading is the clustering coefficient $C$—the fraction of triangles in the network [32]. In general, a high value of this coefficient slows down the spreading [33]. This is quite intuitive. Imagine a triangle, where one node infects the two others. Now the link between the secondary infected nodes is superfluous for the disease spreading and it would have benefitted the spreading if was connected to some distant node instead.

The first quantities investigating the temporal aspects are the average time of the presence of nodes $d_N$ and links $d_L$. To be specific, we define the time of presence as the time between the first and last contact. This will not be perfectly valid, since the last contact does not necessarily mean that a node or link became inactive. However, Ref. 15 indicates that for these data sets the above approximation is not so grave (Ref. 15 studies five datasets in common to this paper).

There has been a good deal of interest in how the distribution of times between contacts affects spreading processes. If this is the only temporal structure present in the data, it is known to slow down epidemic spreading [14]. However, Ref. [15] argues that birth and death of nodes and (more closely related to $d_N$ and $d_L$) are more important for disease-type spreading in empirical data sets.

The final two structural measures try to capture a property that sets *Prostitution* aside from the other data sets, namely that the overall activity is increasing though the sampling period. We measure the fraction of nodes $f_N$ and links $f_L$ that are present in the data set at half the sampling time [34]. In data sets that sample a growing population, one would expect these quantities to be rather low. If the links are stable and the contacts frequent (the time between them short compared to the sampling time), then $f_N$ and $f_L$ are be large.

The results for the above analysis are summarized as radar plots in Fig. 5. We see that *Prostitution* is indeed very different from the others—it has a more heterogeneous degree distribution, it has (as expected) much lower $f_N$ and $f_L$, and it has $C = 0$ (since it is a bipartite network). Among the other networks, *Gallery* is the most special as it has very low $d_N$ and $d_L$ values—not surprising, since it samples gallery visitors coming and going during the sampling period.

**Time evolution of the predicted outbreak diversity**

Next, we look at statistics summarizing histograms like Fig. 4 for all parameter values. We measure the average (Fig. 6) and maximum (Fig. 7) values of the deviation $\Delta\Omega$ of the histograms of the predicted $\Omega$ given the state at $t$. One can think of other summary statistics, but as we will see, we can draw some conclusions that generalize over both the average and maximum. The first observation from Figs. 6 and 7 is that there is more structure in the curves of the temporal networks. This is no surprise since the temporal network and fully mixed cases have a time-invariant overall activity. A second observation is that the decay of the unpredictability (a.k.a. outbreak diversity) $\Delta\Omega$ is not extremely fast for any of the data sets and summary statistics. At $t = 0.2\,T$, i.e. at 20% of the sampling time, $\Delta\Omega$ has rarely decreased to less than 20% of its original value. This should be seen in the context of compartmental models on networks being highly predictable in the sense that outbreaks either die early or converge to deterministic quantities [31]. Our added insight is that even though the latter observation is true, the convergence may be slow. The *Prostitution* data (Figs. 6D and 7D) is a bit different since the values of $\Delta\Omega$ are very low. Probably, the relatively short node and link durations, and the time evolution of the data (reflected in the low *f*-values) accentuate high predictability for the temporal networks further. Furthermore, we see that at $t = 0$ (i.e. with only the seed node known), temporal networks are usually least unpredictable. *Prostitution* is a big exception for the average $\Delta\Omega$ (Fig. 6D) and *Reality* for the maximal $\Omega$ (Fig. 7E).

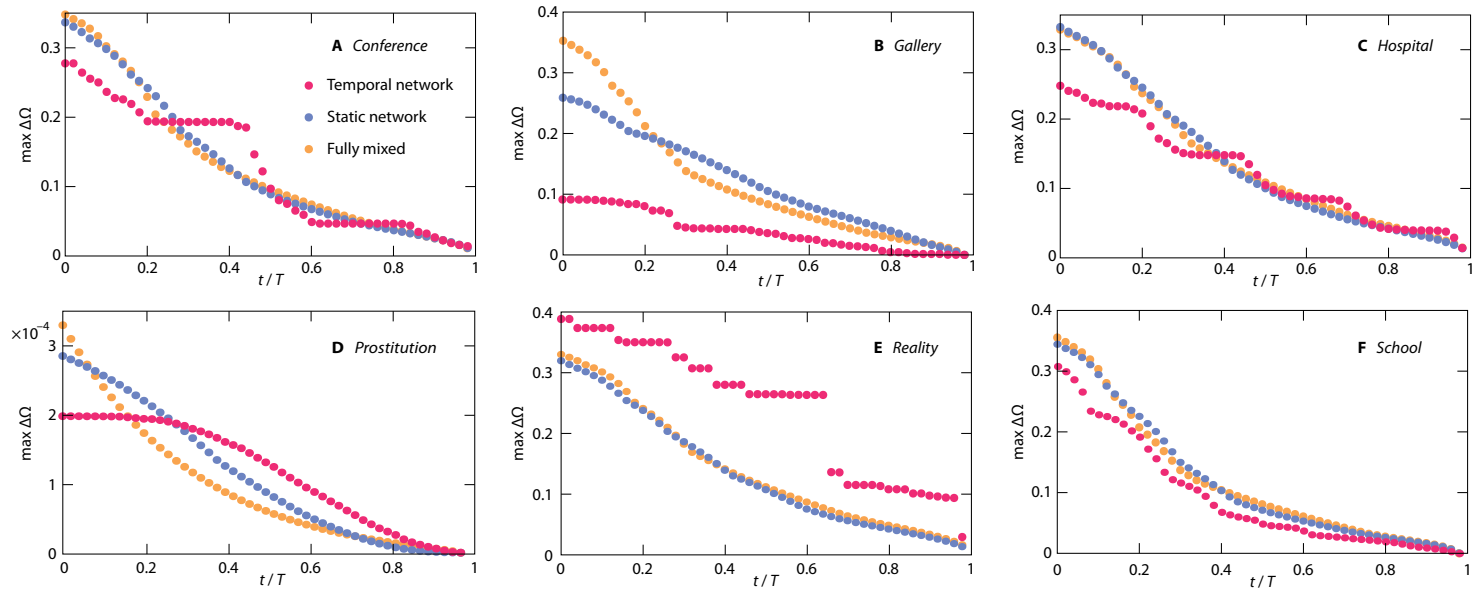

**Fig. 7. The time evolution of the maximum outbreak diversity.** These plots are exactly corresponding to Fig. 4, but for the maximum over the parameter space, rather than the average.

Yet an observation is that the fully mixed case often starts with a higher average (or maximal) $\Delta\Omega$ compared to the static network case, but then decays faster so that for larger $t$ the fully mixed case has smaller outbreak diversity. This tendency is strongest for the networks with most heterogeneous degree distributions (*Prostitution* and *Gallery*). Other than that, it is hard to speculate about the mechanisms for this observation without using a model to tune the network structure (which is an interesting future project, beyond the scope of the present paper).

Our final, and perhaps most interesting observation, is that there is no clear relation between the temporal network representation on one hand and the other two representations on the other hand. For *Gallery* the temporal network representation is more predictable (have smaller outbreak diversity), for *Reality* it is less predictable. We think that small $d_N$ and $d_L$ (like for *Prostitution* and *Gallery*) could, in general, implicate that adding temporal information increases the predictability (as observed above) much. The reason is that then the order of the appearance of nodes and links will matter more. The contacts will then work more like a river system where water flows from higher elevation to lower (or, in our case, from earlier nodes and links to later). Finally, we note that for some data sets (*Conference* and *Prostitution*) the ranking of the representation changes over time. In general, the difference by adding information about time (i.e. going from a static to a temporal network representations) is smaller for $\Delta\Omega$ than $\Omega$ (Fig. 2 and Supplementary Fig. S1).

**Parameter dependence of time to predictability**

Our final analysis regards $\Delta\Omega$'s approach to zero as a function of the SIR parameter values. In other words, we seek to summarize Fig. 3 for $\delta$ and $\lambda$ at the expense of not being able to visualize the full time evolution. Instead we measure the time $t_p$ until there is a 20-fold decrease of the outbreak diversity, i.e. when the deviation of $\Omega$ goes below 0.05 $\Omega_{max}$, where $\Omega_{max}$ is the $\Omega$-value for $\delta = T$ and $\lambda = 1$. The results for the *Gallery* data are plotted in Fig. 8 (for the other data sets—see Supplementary Fig. S2). One interesting observation is that, for all the contact representations, there are parameter values where one has to wait until the very end of the sampling time to get an accurate prediction of the final outbreak size. The inherently hardest prediction happens at long durations and intermediate transmission probabilities. The fact that there is a maximum at intermediate $\lambda$ is probably related to this being the region of longest outbreak times [28]—for smaller $\lambda$, the outbreak dies out quickly; for longer $\lambda$, the outbreak burns out fast in the population. Another reason for the slow approach to predictability is that the outbreaks are less deterministic [31] in this region than for larger $\lambda$ (cf. the discussion of Fig. 4 above). Indeed, a large $\Omega$ does not necessarily mean a short $t_p$. If $\lambda$ is large enough, the stochastic element disappears and the outbreak becomes predictable early (see Fig. 8C).

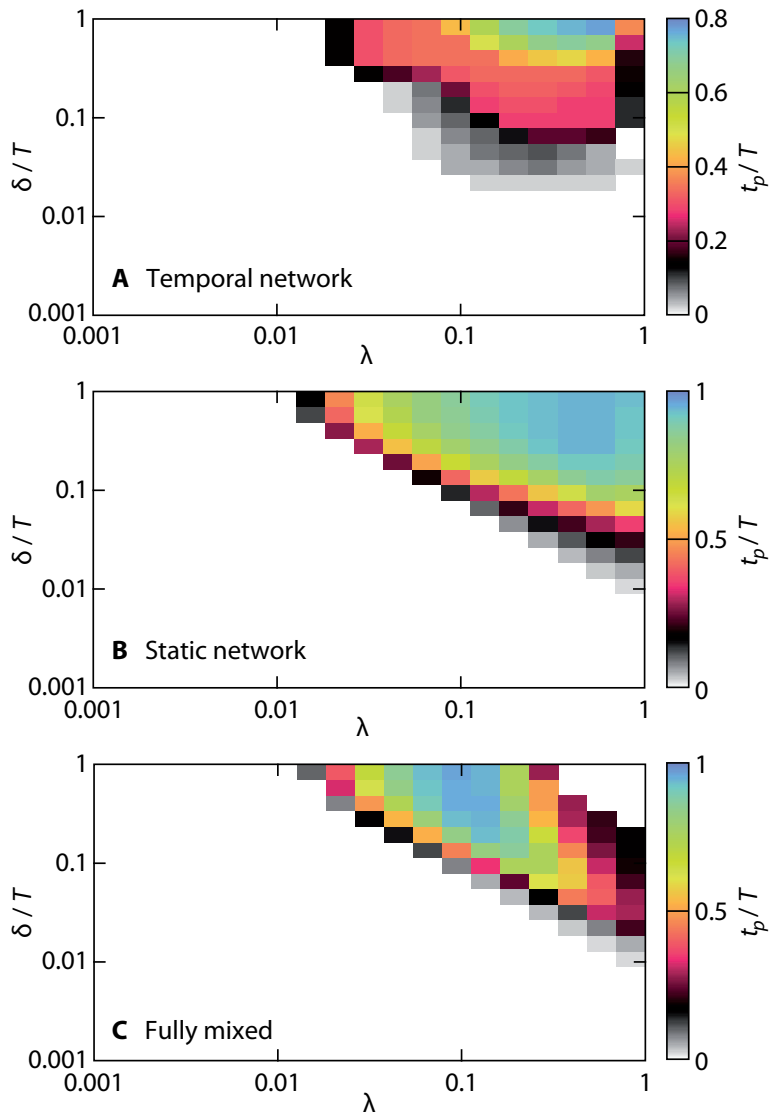

**Fig. 8. Time $t_p$ to reach high predictability.** We define high predictability as when the deviation of the predicted outbreak size is less than 5% of its maximal value. The data set is *Gallery*, these plots for other data sets can be found in Supplementary Fig. S2.

For the two less informative representations, the parameter-space region of slow approach to high predictability is larger. This is true for almost all the datasets (*Prostitution*, once again, being an exception, Supplementary Fig. S2). We also note that, there is more variation in $t_p$ than $\Omega$—for the example *Gallery* data of Fig. 7, all three panels have distinct shapes. For short, the parameter dependence of $t_p$ is more complex than that of $\Omega$. These observations holds for the other data sets with one correction—the densest static networks (*Hospital* and *Gallery*) are very similar to their fully mixed counterparts.

## Discussion

We have studied how the level of information content in the representation of contacts patterns affects the SIR epidemic model. We investigated several aspects of predictability or outbreak diversity—given no knowledge about the outbreak (other than that it happened) and given the state of the system at a breaking time $t$. The starting point of our study was empirical data sets of human proximity. SIR outbreaks in these data sets were mostly slowed down and shrunk when a new layer of information was added (i.e. going from a fully mixed simulation to a static network representation, or going from a static network to a temporal network). Given that we do not know anything about the epidemics (more than it started), a classic (differential-equation based) analysis would overestimate the severity of the disease, as would a static-network based model. On the other hand, if we instead study the histogram of future outbreak sizes given the state of the system at time $t$, then there is no clear trend with respect on the information content (still, the deviations can be large). In other words, different representations do give different results, but it is, strictly speaking, not the case that adding information systematically increases or decreases the deviation of the predicted outbreak sizes. This itself seems to depend on the temporal network structure in a complex way, that this paper only takes a first step towards understanding.

It is hard to generalize all features of the outbreak diversity. We note that for most data sets, including more information about the contacts makes the outbreaks smaller. However, this is not always the case (as the *Prostitution* data behaves the other way around [12,14]). Another fairly universal feature is that, for later times (initially it could be the other way around) the fully connected topologies are more predictable than the static networks. On the other hand, outbreaks on the temporal networks can be both more or less predictable. We note that the data sets with relatively short durations of the presence of nodes and links (the time between the first and last time they are observed) lose most predictability by projecting out the temporal information.

Not all our observations are vague—we see can clearly see the importance of the temporal structures. Going from a temporal to a static representation can quantitatively make a big difference, but not only that—the outbreak distributions are either bi- or unimodal for the static network (and fully mixed) simulations, whereas for the temporal networks, the distribution can be multimodal (cf. Fig. 3).

This work is a starting point. To corroborate the observations above one approach would be to use generative models that can tune the temporal network structure of the data (cf. Refs [10,13,30]). In general, we anticipate much research relating aspects of the available information about an epidemic outbreak, and the contact structure of the population, with the predictability of the outbreak.

## Methods

In this section, we will go through technicalities of the SIR model that are not fully explained above.

|  | Number of individuals | Number of contacts | Sampling time | Time resolution |
| --- | --- | --- | --- | --- |
| *Conference* | 113 | 20,818 | 2.50d | 20s |
| *Gallery* | 200 | 5,943 | 7.80h | 20s |
| *Prostitution* | 16,730 | 50,632 | 6.00y | 1d |
| *Hospital* | 75 | 32,424 | 96.5h | 20s |
| *Reality mining* | 64 | 26,260 | 8.63h | 5s |
| *School* | 236 | 37402 | 8.61h | 20s |

**Table 1. The basic statistics of the data sets.**

**SIR simulations**

In this work, we use the constant duration version of the SIR model [35]. We initialize all nodes to susceptible except pick one random individual *i* that we set to infectious. We assume that *i* becomes infectious at the same time as its first appearance in the data (i.e. it can infect others starting from its first contact). In a contact between an infectious and susceptible, the susceptible will (instantaneously) become infectious with a probability λ. Infectious individuals stay infectious for δ time steps whereupon they become recovered. If more than one contact occur during the same time step, we go through them in a random order.

Another common version of the SIR model is to let infectious individuals recover with a constant rate. Qualitatively, both versions give the same results [35]. We use the constant duration version because it both has a peaked distribution of the infection times (as opposed to the exponentially distributed times of the constant recovery rate version) and makes the code a bit faster than the exponentially distributed durations.

For all parameter values, all data sets and all representations, the output is averaged over $10^3$ independent outbreaks (and every time step of every outbreak is the staring point of $10^4$ auxiliary runs, as mentioned above).